\documentclass[]{interact}

\usepackage[utf8]{inputenc}
\usepackage[T1]{fontenc}
\usepackage{epstopdf}
\usepackage{subfigure}
\usepackage{multirow}
\usepackage{colortbl}
\usepackage{xcolor}
\usepackage{hhline}
\usepackage{url}

\usepackage{natbib}
\bibpunct[, ]{(}{)}{;}{a}{}{,}

\definecolor{darkyellow}{RGB}{230, 215, 0}

\begin{document}

\articletype{ARTICLE}

\title{A case study of hybrid statistical models for analyzing spatio-temporal effects in historical map series of Donau-Ries county}

\author{
\name{Mireille Fangueng\textsuperscript{a}\thanks{CONTACT Mireille Fangueng. Email: Mireille.Fangueng@ikg.uni-hannover.de}, Frank Thiemann\textsuperscript{a}, Monika Sester\textsuperscript{a} and Philipp Otto\textsuperscript{b}}
\affil{\textsuperscript{a}Institute of Cartography and Geoinformatics, Leibniz University Hannover, Appelstra\ss e 9a, 30167 Hannover, Lower Saxony, Germany; \textsuperscript{b}School of Mathematics and Statistics, University of Glasgow, University Place, Glasgow G12 8QQ, Scotland, United Kingdom}
}

\maketitle

\begin{abstract}
This case study compares generalized linear models, generalized additive models, and semi-structured deep distributional regression for analyzing settlement patterns in historical land-cover maps. Using four reference years for Donau-Ries county (Landkreis Donau-Ries) in southern Germany, we quantify how distances to agricultural areas, forests, and rivers relate to residential presence, and we evaluate predictive performance through spatial block cross-validation. Classical models yield stable and interpretable distance effects, whereas the semi-structured framework improves prediction once local image context is included. The analysis illustrates a practical trade-off between interpretability, accuracy, and robustness when interpretable covariates are combined with deep components in spatio-temporal data analysis.
\end{abstract}


\vspace{6pt}
\noindent\textit{Approximate word count of the main text is 9{,}900 words, excluding tables, figures and references.}

% =============================================================================
\section{Introduction}\label{sec1}

The analysis of land use dynamics using historical spatial data plays a crucial role in understanding long-term urbanization processes and their relationship with landscape structure. In particular, the spatial distribution of settlement areas is associated with proximity to key environmental features such as agricultural areas, forests, and rivers, which shape accessibility, land availability, environmental constraints, and socio-economic attractiveness. Understanding how these spatial factors relate to urban presence and how their estimated effects evolve over time remains a central challenge in land use analysis. Recent work \citep{aslam2023monitoring, hu2024five} reflects the growing interest in combining multi-temporal spatial data with statistical and machine learning approaches to characterize urban dynamics.

The three landscape distances used in this article are interpretable proxies for landscape structure rather than an exhaustive description of the forces that produce urban growth. Settlement dynamics also respond to demographic pressure, planning regulation, transport accessibility, and land markets, and none of these dimensions is recorded in the historical map series analyzed here. The effects reported in the following sections are therefore conditional associations estimated within a fixed covariate set, and they are not intended as causal statements about urban expansion.

From a methodological perspective, this challenge is twofold. First, models must be capable of capturing non-linear and spatially heterogeneous relationships between urbanization and landscape features. Second, they must remain interpretable, allowing meaningful insights into the spatial regularities behind settlement patterns rather than focusing solely on predictive performance. In this context, interpretable statistical models offer a valuable framework for describing settlement dynamics while ensuring transparency and scientific interpretability.

Generalized linear models (GLMs; \citealt{nelder1972generalized}) have long been used to model binary land use outcomes, such as the presence or absence of urban areas, by relating them to explanatory variables through linear predictors. When applied to settlement modeling, GLMs allow the estimation of odds ratios that directly quantify how distances to environmental features are associated with the likelihood of urban presence. However, the assumption of linear relationships may limit their ability to represent the complex spatial processes underlying land conversion.

Generalized additive models (GAMs; \citealt{hastie1985generalized}) extend GLMs by introducing smooth functions that capture non-linear effects of predictors. This flexibility makes GAMs particularly suitable for modeling distance-based relationships, where urbanization probability may vary in a non-monotonic manner with distance to agricultural areas, forests, or rivers. GAMs have therefore been widely used to explore spatial gradients and threshold effects in land use change, while preserving a high level of interpretability through smooth terms \citep{feng2017gamca}.

More recently, hybrid modeling approaches have emerged to bridge the gap between interpretable statistical models and data-driven deep learning methods. Several families of approaches are relevant in this context. Geographically weighted artificial neural network regression (GWANN; \citealt{hagenauer2022geographically}) and geographically neural network weighted regression (GNNWR; \citealt{du2020geographically}) extend classical geographically weighted regression by replacing the local linear estimator with a neural network, allowing the simultaneous modeling of spatial heterogeneity and non-linear predictor effects. These approaches have shown promise in spatial estimation tasks where effects vary substantially across the study area. Gradient boosting methods, such as XGBoost or LightGBM, combined with spatial cross-validation and SHAP-based interpretability \citep{lundberg2017unified}, represent another family of hybrid approaches that offer high predictive performance while enabling post-hoc decomposition of variable contributions. Such methods have been applied to land use classification and urban suitability mapping \citep[e.g.][]{hu2024five}. Semi-structured deep distributional regression (SDDR), implemented in the PySDDR framework \citep{rugamer2023deepregression, rugamer2024semi}, combines structured additive predictors such as spline-based smooth effects with unstructured deep neural networks. This architecture allows the explicit modeling of interpretable distance effects while simultaneously capturing higher-order spatial dependencies that may not be fully described by traditional models. As such, SDDR provides a promising framework for analyzing settlement patterns in complex spatial environments.

Beyond model structure, a key challenge in land use modeling lies in accounting for temporal dependence. Historical maps, aerial imagery, and satellite data provide valuable information on past landscape configurations, which may inform present-day urbanization patterns. While multi-temporal and spatiotemporal approaches have been widely explored in remote sensing and landscape ecology \citep[e.g.][]{haase2007changes, turek2018example}, the explicit integration of temporal context into interpretable hybrid models remains relatively underexplored. In particular, few studies have examined how incorporating historical imagery affects both the estimated spatial effects of environmental variables and the predictive performance of urbanization models. Some studies have explored the use of historical aerial imagery or historical maps to document landscape evolution. For example, \cite{nocerino2012multi} introduced a methodology using historical aerial imagery to construct spatio-temporal representations of landscapes by combining photo-interpretation and 3D reconstruction techniques. Similarly, \cite{nebiker2014building} demonstrated how historical aerial photographs can be used to detect long-term changes in settlement structures through dense image matching and object-based image analysis. Other studies have combined historical maps and aerial imagery to analyze long-term landscape dynamics using machine learning approaches \citep{liu2018integration}. While these works highlight the value of historical geospatial data for reconstructing past landscapes, they generally focus on change detection or landscape reconstruction rather than on integrating historical imagery into statistical models that describe or predict urbanization dynamics. In particular, relatively little attention has been paid to how historical spatial data derived from such imagery can be used within an explicit statistical comparison of modeling frameworks.

Building on this gap, the present article is a comparative case study of GLMs, GAMs, and SDDR models applied to settlement presence over multiple time periods. Our objective is to assess how each modeling framework represents the spatial and temporal effects of distances to key landscape elements, and to evaluate the trade-offs between interpretability and predictive performance. By analyzing model coefficients, smooth terms, and partial effects across time, we aim to provide a coherent and interpretable description of settlement dynamics, while identifying the modeling approach best suited to long-term analyses of this type of data. The contribution of the article is therefore methodological and empirical rather than substantive, since the comparison is conducted under one fixed covariate set and one study region.

Donau-Ries county (Landkreis Donau-Ries) in Bavaria, southern Germany, is used as a case study because of the availability of high-resolution historical land cover data across multiple time points, namely 2004, 2009, 2015, and 2023. Rather than being the primary object of inference, this region provides a realistic and challenging environmental setting to examine how different statistical modeling strategies handle spatial nonlinearity, temporal evolution, and contextual information. In particular, we investigate the added value of incorporating temporal image context into SDDR models and analyze how this affects both predictive performance and the estimated partial effects of environmental distances.

The remainder of the article is organized as follows. Section~\ref{sec2} describes the study area and data preprocessing and presents the GLM, GAM, and SDDR modeling frameworks together with their implementation. Section~\ref{sec3} reports the results, including the interpretation of GLM coefficients, the temporal analysis of GAM smooth terms, and the partial effects derived from SDDR models, followed by a comparative evaluation of predictive performance in a temporal context. Section~\ref{sec4} provides a critical discussion of the results, with particular emphasis on methodological trade-offs, model interpretability, and robustness in a spatio-temporal setting. Finally, Section~\ref{sec5} summarizes the main findings and outlines perspectives for future research on interpretable modeling of land use change.

% =============================================================================
\section{Materials and methods}\label{sec2}

\subsection{Study area and data}\label{subsec:area}

\subsubsection{Study area}

\begin{figure}[h!]
\centering
\includegraphics[scale=0.08]{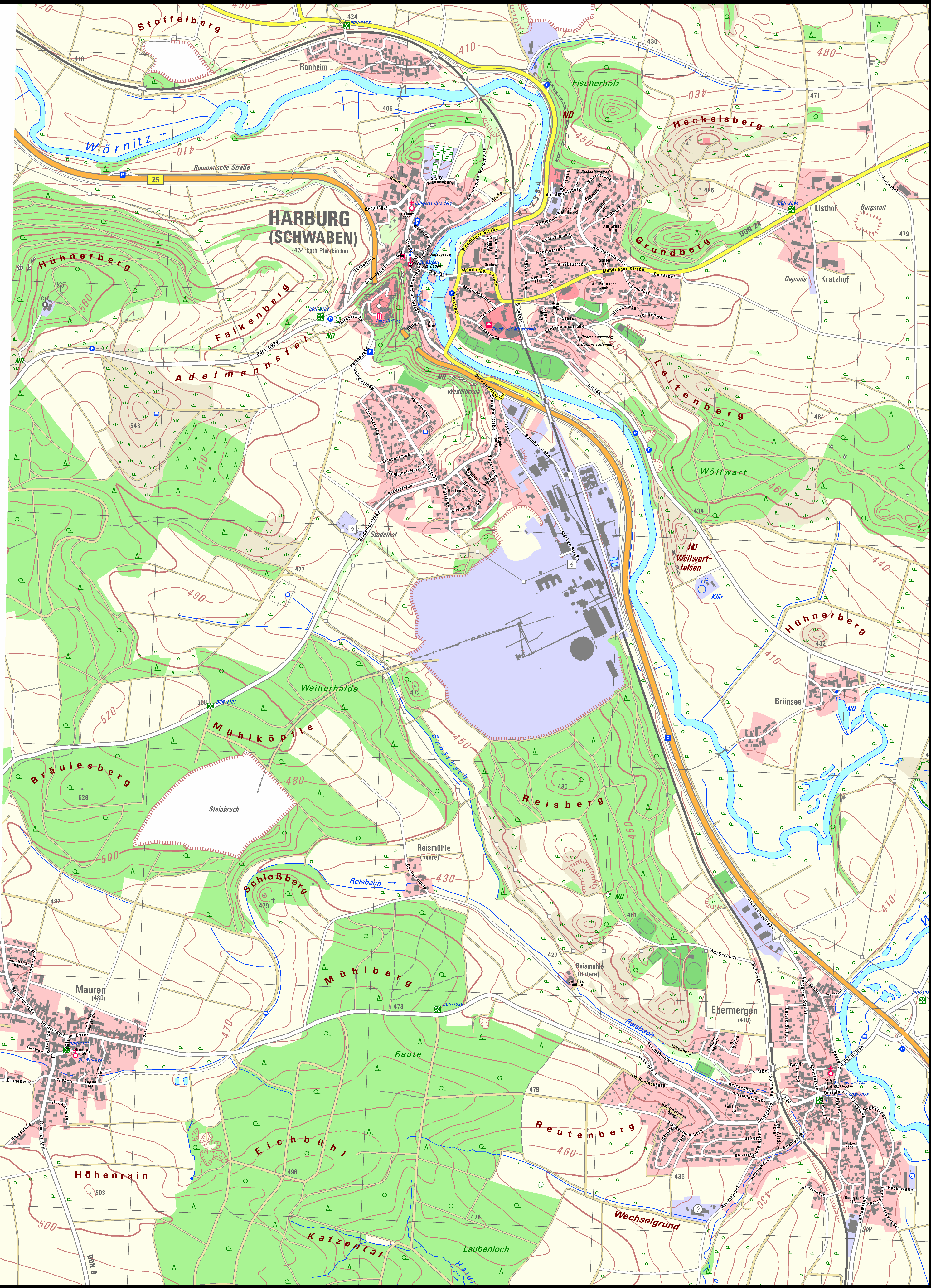}
\caption{Land-cover map of Donau-Ries county from 2015. Reproduced from TK25 sheet 7230, \copyright\ Bayerische Vermessungsverwaltung 2015.}
\label{fig:map2015}
\end{figure}

The study area is Donau-Ries county (Landkreis Donau-Ries) in Bavaria, southern Germany, which was selected because of its heterogeneous landscape structure. This region exhibits a diversified spatial configuration, combining agricultural areas, forested zones, river networks, and residential settlements, making it particularly suitable for investigating the spatial correlates of urban presence (see Figure~\ref{fig:map2015}). The county is characterized by a close interweaving of natural and anthropogenic land cover types, where settlement areas have historically developed in interaction with agricultural production zones, forest margins, and fluvial systems. In this study, land cover maps were analyzed for four reference years, namely 2004, 2009, 2015, and 2023, enabling the assessment of urban expansion and land use dynamics over nearly two decades. This temporal depth allows the evaluation of temporal dependencies in the relationship between urbanization and surrounding landscape features.

\subsubsection{Data preprocessing}\label{subsec:preproc}

\begin{figure}[h!]
\centering
\includegraphics[scale=0.5]{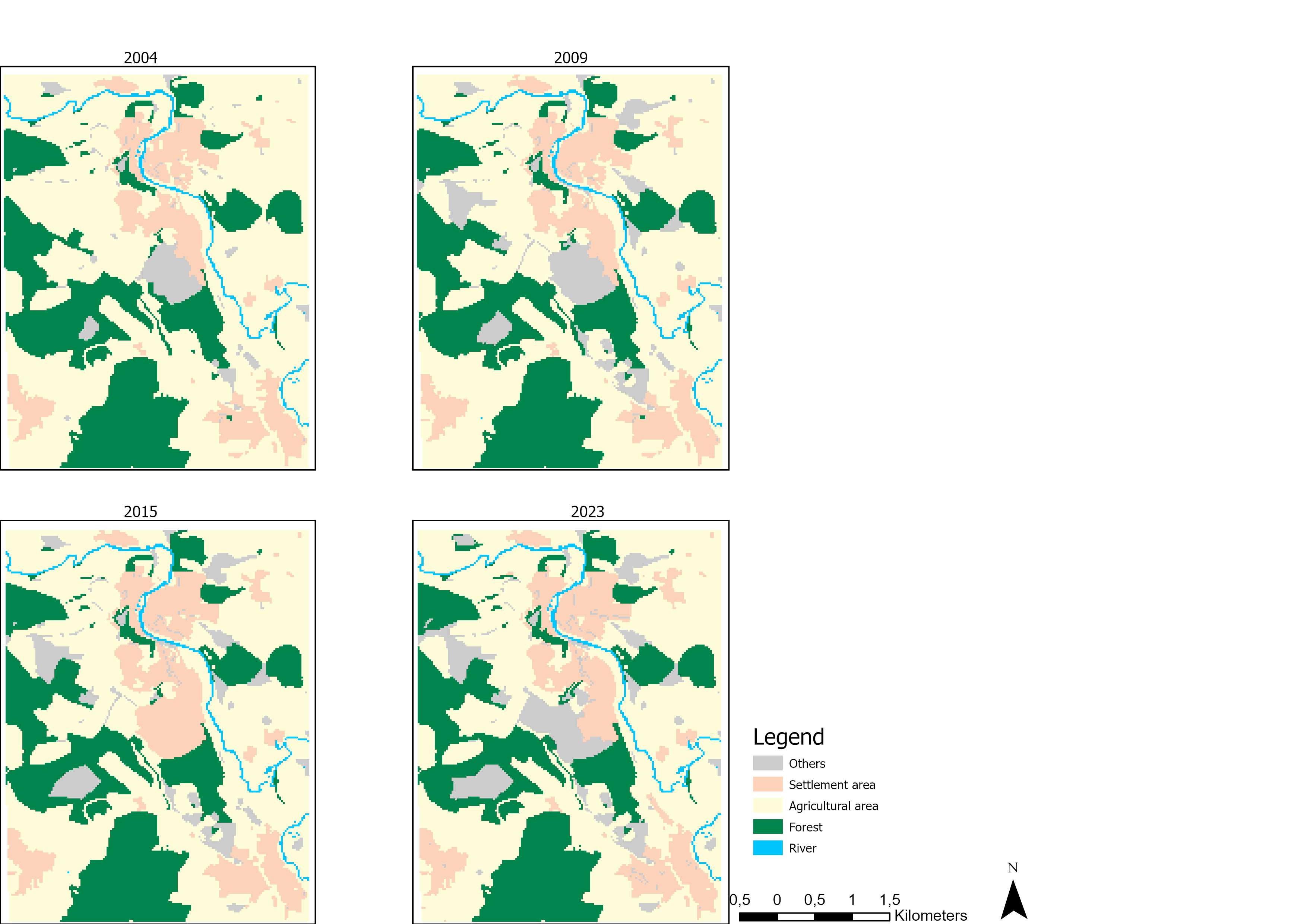}
\caption{Spatio-temporal evolution of land use in the study area between 2004 and 2023. The maps show the spatial distribution of forest, agricultural areas, settlement areas, rivers and other land cover classes for four reference years. All maps share the same spatial extent, projection, scale and symbology to ensure direct visual comparison of land use changes over time. Visual inspection suggests a progressive expansion of settlement areas, particularly along the central valley, accompanied by local transformations of surrounding agricultural land. These visible patterns motivate the quantitative modeling of settlement dynamics developed in the following sections.}
\label{fig:maps2004_2023}
\end{figure}

Land cover data for Donau-Ries county (see Figure~\ref{fig:maps2004_2023}) were processed using ArcGIS software. The original land cover maps, represented as two-dimensional raster matrices whose elements correspond to pixels, were first transformed into a one-dimensional column vector to enable pixel-wise statistical modeling. In this representation, each pixel constitutes a single observation. Specific layers representing settlement area (AX52001, AX41001 and AX41006), agricultural areas (AX43001), forests (AX54001 and AX43002), and rivers (AX44001) were extracted from the ATKIS Digital Landscape Model (DLM) base dataset and subsequently rasterized onto a $25$-meter grid, ensuring a consistent spatial resolution across all variables.

The dependent variable $Y_{i,t}$ indicates whether an area is urban (settlement). It was obtained by vectorizing the raster map of settlement areas into a binary column vector, where $Y_{i,t} = 1$ if the $i$-th pixel at time $t$ belongs to a residential zone, and $Y_{i,t} = 0$ otherwise. This transformation allows the spatial distribution of settlement areas to be analyzed using standard regression-based frameworks while preserving the underlying spatial structure through pixel indexing.

To characterize the spatial correlates of urbanization, we computed isotropic Euclidean distances from each pixel to the nearest agricultural area, forest, and river. These distance-based predictors were selected because they represent fundamental and relatively stable components of the landscape that are strongly associated with settlement patterns through a combination of accessibility, environmental constraints, and amenity effects. Agricultural areas typically correspond to open and easily convertible spaces that favor urban expansion, forests introduce ecological and regulatory constraints while also providing landscape amenities at intermediate distances, and rivers act as major structuring elements, offering both attractive features and environmental risks.

Distances were computed such that pixels located within a given land cover class were assigned a distance value of zero. For instance, the distance to agricultural areas measures the proximity of each pixel to the nearest agricultural zone, with a value of zero indicating pixels already classified as agricultural. This formulation allows the construction of continuous spatial gradients that are well suited for modeling non-linear effects using GAM and SDDR frameworks, while remaining directly interpretable within a GLM context.

The choice of these predictors reflects a deliberate trade-off between explanatory power and interpretability. By focusing on distances to agricultural areas, forests, and rivers, the analysis isolates a small set of landscape gradients that can be estimated consistently and compared directly across the different modeling approaches.

This restricted covariate set is a deliberate scope limitation of the study and should be read as such. Urban expansion is also governed by population growth, employment structure, land prices, zoning decisions, and infrastructure investment, and these socio-economic drivers are not available for the four reference years considered here. The three landscape distances therefore act as observable proxies for accessibility, land availability, and environmental constraint rather than as a complete explanation of settlement formation. This is also why the unstructured component of the SDDR framework matters for the present comparison, since it absorbs part of the spatial signal that the three distances leave unexplained.

\subsection{Statistical modeling framework}\label{subsec:framework}

To explore the correlates of residential area presence, distances to key land use categories, namely agricultural area, forests, and rivers, were computed and used as explanatory variables. The overarching objective of the modeling framework is to identify the most influential predictors and to quantify how these spatial factors are associated with the probability of residential presence. We adopt a progressive modeling strategy, starting with simple and interpretable statistical models and gradually introducing greater flexibility to better account for the spatial complexity inherent in geographic data. In particular, this approach allows us to assess how much explanatory power is gained when moving from strictly linear assumptions to models capable of representing non-linear and spatially heterogeneous relationships.

\subsubsection{Generalized linear model}

The GLM was used as a baseline approach to describe the relationship between residential presence and distance-based predictors. GLMs extend classical linear regression to response variables that follow distributions from the exponential family, making them well suited for environmental and land use studies involving non-Gaussian outcomes. The GLM was implemented in \textsf{R} using the \texttt{stats} package, which provides maximum likelihood estimation within the classical regression framework.

In this study, the response variable is binary and indicates whether a given pixel corresponds to a residential area or not. We therefore assume a binomial distribution, where each pixel represents a single trial with two possible outcomes, namely success or failure. Here, a success corresponds to the presence of a residential area at pixel $i$ and time $t$, while a failure corresponds to its absence. Formally, the response variable $Y_{i,t}$ is defined as
\begin{equation}
    Y_{i,t} \sim \mathcal{B}(n, p_{i,t}),
\end{equation}
where \( Y_{i,t} = 1 \) indicates that the \( i \)-th pixel at time $t$ is classified as residential, and \( Y_{i,t} = 0 \) otherwise. Since each pixel represents a single observation, \(n=1\), and the model estimates the probability $p_{i,t}$ of observing a residential area at that location and time. The GLM links this probability to the explanatory variables using a logit link function, which maps probabilities bounded between 0 and 1 onto the real line,
\begin{equation}
    g(E(Y_{i,t})) = \log\left(\frac{p_{i,t}}{1 - p_{i,t}}\right) = \beta_0 + \beta_1 X_{1i,t} + \beta_2 X_{2i,t} + \beta_3 X_{3i,t},
\end{equation}
where \( X_{1i,t}, X_{2i,t}, X_{3i,t} \) represent the distances to agricultural area, forest areas, and rivers, respectively, for pixel $i$ at time $t$. This formulation allows us to examine the direct linear effects of each explanatory variable on the probability of a residential area in each pixel. The GLM was selected because it provides a straightforward and interpretable set of estimated coefficients. For example, a positive coefficient $\beta_k$ indicates that an increase in the associated explanatory variable $X_{k}$ increases the probability of the pixel being a residential area. While the GLM is robust for capturing linear relationships, it may not be suitable for modeling more complex, non-linear relationships between the explanatory and response variables. This limitation motivated the consideration of a more flexible GAM.

\subsubsection{Generalized additive model}

To better capture potential non-linear relationships between the distances to various geographic features and the presence of residential areas, we applied a GAM. It was fitted in \textsf{R} using the \texttt{mgcv} package \citep{wood2017generalized}, which allows flexible estimation of smooth effects within a penalized regression framework. Unlike the GLM, the GAM allows for greater flexibility by modeling some relationships as non-linear through smooth functions. It can be expressed as follows,
\begin{equation}
    \begin{aligned}
        g(E(Y_{i,t})) &=
        \log\left(\frac{p_{i,t}}{1 - p_{i,t}}\right) \\ &=
        \beta_0 + f_{1,t}(X_{1i,t}) + f_{2,t}(X_{2i,t})
        + f_{3,t}(X_{3i,t}) + f_{4,t}(\text{long}_{i,t}, \text{lat}_{i,t}),
    \end{aligned}
\end{equation}
where \( f_{1,t}(\cdot) \), \( f_{2,t}(\cdot) \), and \( f_{3,t}(\cdot) \) are smooth functions for the distance variables, and \( f_{4,t}(\cdot, \cdot )\) is a two-dimensional smooth spatial term.

The smooth terms were estimated using penalized splines, which help control the complexity of the smoothness and prevent overfitting. The use of penalized splines is critical in the GAM \citep{eilers1996flexible}, as it allows for the estimation of non-linear effects while avoiding excessive flexibility. This is particularly useful for geographic distance variables, where the relationship with the probability of a residential area may not be strictly linear. For example, the effect of distance to rivers may decay non-linearly as the distance increases. Additionally, the inclusion of the two-dimensional smooth spatial term $f_{4,t}(\text{long}_{i,t}, \text{lat}_{i,t})$ allows us to model complex spatial variations that cannot be captured by the distance variables alone. This term accounts for potential spatial effects due to unobserved geographic factors such as population density or local infrastructure. Compared to the GLM, the GAM offers a more flexible and potentially more accurate approach to modeling spatially complex phenomena. However, when spatial context becomes high-dimensional, for example when incorporating image-based information, the semi-parametric structure of GAMs may reach its limits. In this setting, the SDDR framework extends the GAM by embedding neural network components within an interpretable structured predictor, thereby combining statistical inference with deep representation learning.

\subsubsection{Semi-structured deep distributional regression}

To further investigate the spatial correlates of residential area presence, we employ the SDDR framework implemented in the PySDDR package, the Python counterpart of the \textit{deepregression} framework \citep{rugamer2023deepregression}. This modeling approach extends generalized additive models by combining structured additive predictors with deep neural networks, thereby enabling the joint use of heterogeneous data sources within a single coherent model. In our application, the structured component of the SDDR model relies on tabular data derived from the spatial configuration of the landscape. Specifically, the input table consists of distance-based predictors, where each column corresponds to the distance from a given pixel to a key land use category, namely agricultural area, forest areas, and rivers, together with the response variable. This structure is similar to that used in the GLM and GAM frameworks, ensuring continuity and comparability across modeling approaches. Beyond these structured predictors, SDDR allows for the incorporation of unstructured spatial information through a neural network component. The model formulation is given by
\begin{equation}
    \begin{aligned}
        g(E(Y_{i,t})) &=
        \log\left(\frac{p_{i,t}}{1 - p_{i,t}}\right) \\ &=
        \beta_0 + f_{1,t}(X_{1i,t}) + f_{2,t}(X_{2i,t})
        + f_{3,t}(X_{3i,t}) + ANN(X_{i,t}),
    \end{aligned}
    \label{equation4}
\end{equation}
where \( f_{1,t}(\cdot) \), \( f_{2,t}(\cdot) \), and \( f_{3,t}(\cdot) \) are smooth functions capturing non-linear effects of distances to agricultural area, forests, and rivers, respectively. These components retain the interpretability of GAMs by explicitly modeling how each distance variable is associated with the probability of residential presence. The term \( ANN(X_{i,t}) \) represents the unstructured deep learning component of the model. Rather than relying solely on scalar distance values, this component exploits local spatial context by using image-based information. For each pixel, a sub-image centered on that pixel is extracted from the larger spatial dataset, defining a neighborhood whose size is controlled by a predefined radius. In this study, neighborhoods corresponding to radii of 2, 8 and 16 pixels were considered, allowing the model to capture spatial patterns at multiple scales. These sub-images encode local land use configurations and spatial textures that cannot be fully represented by distance variables alone. The artificial neural network processes this information to learn complex spatial interactions and higher-order dependencies relevant to urbanization processes.

By combining structured smooth effects with an unstructured neural network component, the SDDR framework offers a flexible yet interpretable approach to modeling land use dynamics. The structured terms provide direct insights into the marginal effects of key spatial predictors, while the neural network captures residual spatial information related to local context and neighborhood structure. This hybrid formulation allows the model to better account for spatial heterogeneity and non-linear interactions, thereby extending the capabilities of both GLM and GAM approaches. To assess the impact of model complexity and spatial context, three different SDDR specifications were explored, each corresponding to a distinct architecture and neighborhood configuration. These variants allow us to evaluate how the inclusion and scale of image-based contextual information influence predictive performance and model behavior in a temporal urbanization setting.

The neural network architectures used within the SDDR framework were deliberately kept simple and closely aligned with the reference implementations provided in the exploratory version of the \texttt{PySDDR} package. Rather than pursuing an extensive hyperparameter optimization or architecture search, our objective was to assess the added value of integrating deep learning components into structured additive models under a controlled and interpretable setting.

For Models~1 and~2, which rely exclusively on tabular distance-based predictors, the neural components were designed to operate at the level of individual covariates. In Model~1, each distance variable is associated with a single-layer network of the form \texttt{Linear(1,15)}. This architecture corresponds to a linear projection of a univariate input into a higher-dimensional latent space, allowing the model to learn multiple weighted representations of the same predictor while preserving a strong link with the original variable. Such a design introduces additional flexibility beyond classical smooth terms, without obscuring interpretability or inducing complex interactions that would be difficult to attribute to specific predictors.

Model~2 extends this principle by introducing shallow multi-layer perceptrons of the form \texttt{Linear(1,3) $\rightarrow$ ReLU $\rightarrow$ Linear(3,8)} for each distance variable. This architecture enables the learning of non-linear feature transformations while remaining intentionally shallow. The limited depth and small number of hidden units were chosen to balance expressive power and model stability, ensuring that the neural components complement rather than dominate the structured spline-based effects. Importantly, this configuration allows the assessment of whether modest increases in network depth lead to measurable gains in predictive performance, without departing from an interpretable semi-parametric framework.

For Model~3, which incorporates unstructured spatial information, a dedicated neural network was used to process image-based neighborhoods centered on each pixel. The architecture consists of a \texttt{Flatten(1, -1)} layer followed by a fully connected layer \texttt{Linear($p \times p$, $a$)}\footnote{Here $p= 2r+1$ and $a = 2p$, where $r$ is the radius defined for the different neighborhoods around each pixel in the construction of sub-images.} and a ReLU activation function. In the case of $5 \times 5$ pixel windows, this corresponds to \texttt{Linear(25,10)}. This design reflects a conscious modeling choice, since the goal is not to perform deep convolutional feature extraction, but rather to summarize local spatial context through a compact, learnable representation that can be seamlessly integrated into the structured additive predictor. By avoiding deeper convolutional architectures, the model maintains computational efficiency and reduces the risk of overfitting, which is particularly important given the moderate size of historical land-use datasets.

Overall, the selected architectures prioritize transparency, parsimony, and methodological consistency across models. By adopting network structures that are simple, well defined, and aligned with the default configurations of the \texttt{PySDDR} framework, this study focuses on evaluating the conceptual contribution of neural components within SDDR rather than optimizing deep learning performance in isolation. This strategy ensures that any observed performance improvements can be attributed to the integration of spatial and temporal information, rather than to aggressive architectural tuning.

\begin{itemize}
    \item \textit{Model 1, structured deep regression with single-layer networks.}

This first SDDR specification closely follows the structure of a classical structured additive regression model, while introducing a minimal deep learning component to enrich the representation of distance-based effects. The model combines smooth spline terms with simple neural network transformations applied independently to each distance predictor, thereby preserving interpretability while allowing for additional flexibility. The structured component of the model consists of smooth spline functions applied to the distances to agricultural areas, forest areas, and rivers. These terms, denoted \( f_{1,t}(X_{1i,t}) \), \( f_{2,t}(X_{2i,t}) \), and \( f_{3,t}(X_{3i,t}) \), capture potentially non-linear but interpretable relationships between each distance variable and the probability of residential presence. As in GAMs, these smooth effects represent the dominant, explainable contribution of each predictor. In parallel, each distance variable is also passed through a dedicated neural network component, denoted \( ANN_{1,t}(X_{1i,t}) \), \( ANN_{2,t}(X_{2i,t}) \), and \( ANN_{3,t}(X_{3i,t}) \), respectively. Importantly, each neural network operates on a single predictor only, ensuring that no interactions between predictors are introduced at this stage. The resulting model is expressed as
\begin{equation}
    \begin{aligned}
        g(E(Y_{i,t})) &=
        \log\left(\frac{p_{i,t}}{1 - p_{i,t}}\right) \\ &=
        \beta_0
        + f_{1,t}(X_{1i,t})
        + f_{2,t}(X_{2i,t})
        + f_{3,t}(X_{3i,t}) \\
        &\quad + ANN_{1,t}(X_{1i,t})
        + ANN_{2,t}(X_{2i,t})
        + ANN_{3,t}(X_{3i,t}),
    \end{aligned}
\end{equation}
where \( X_{1i,t} \), \( X_{2i,t} \), and \( X_{3i,t} \) denote the distances to agricultural area, forest, and river, respectively. Each neural network component is implemented as a single fully connected layer defined by \texttt{Linear(1,15)}. The first dimension indicates that the input to the network is one-dimensional, corresponding to a single scalar distance value for a given pixel. The second dimension represents the number of hidden units in the layer, meaning that the scalar input is projected into a 15-dimensional latent feature space. Formally, each component can be written as
\[
ANN_{j,t}(X_{ji,t}) = W_j X_{ji,t} + b_j,
\]
where \( W_j \in \mathbb{R}^{15 \times 1} \) is a weight matrix and \( b_j \in \mathbb{R}^{15} \) is a bias vector. These neural transformations do not introduce non-linearity through activation functions. Instead, they act as learnable linear embeddings of the distance variables. Their role is to allow the model to re-weight and re-express each predictor in a higher-dimensional space, thereby complementing the smooth spline terms without compromising interpretability. In this sense, the ANN components can be viewed as flexible, data-driven extensions of linear effects that coexist with the explicitly modeled smooth functions. Overall, this first SDDR specification represents a conservative hybrid model, in which the spline-based structured terms remain the primary interpretable drivers of the predictions, while the single-layer neural networks provide a limited yet effective enhancement of the expressive power of the model. This design makes Model~1 particularly suitable as a baseline hybrid approach bridging classical GAMs and more complex deep regression architectures.

\item \textit{Model 2, structured deep regression with multi-layer networks.}

The second SDDR specification extends Model~1 by increasing the depth and expressiveness of the neural network components associated with each distance-based predictor. While the structured part of the model remains unchanged and continues to rely on smooth spline functions for interpretability, the deep component is enriched through multi-layer neural networks, allowing the model to capture more complex and potentially non-linear transformations of the distance variables. As in the previous specification, the structured component consists of smooth functions applied to the distances to agricultural area, forest, and river, denoted \( f_{1,t}(X_{1i,t}) \), \( f_{2,t}(X_{2i,t}) \), and \( f_{3,t}(X_{3i,t}) \). These terms represent the primary interpretable effects and describe how the probability of residential presence varies smoothly with distance to each land-use category. The key difference with Model~1 lies in the architecture of the neural network components, denoted here as \( ANN_{1,t}(X_{1i,t}) \), \( ANN_{2,t}(X_{2i,t} )\), and \( ANN_{3,t}(X_{3i,t}) \). Each of these components is implemented as a multi-layer feedforward neural network composed of two linear transformations separated by a non-linear activation function. This design allows each predictor to be mapped into a richer latent feature space before contributing to the final predictor. The resulting model is expressed as
\begin{equation}
    \begin{aligned}
        g(E(Y_{i,t})) &=
        \log\left(\frac{p_{i,t}}{1 - p_{i,t}}\right) \\ &=
        \beta_0
        + f_{1,t}(X_{1i,t})
        + f_{2,t}(X_{2i,t})
        + f_{3,t}(X_{3i,t}) \\
        &\quad + ANN_{1,t}(X_{1i,t})
        + ANN_{2,t}(X_{2i,t})
        + ANN_{3,t}(X_{3i,t}),
    \end{aligned}
\end{equation}
where \( X_{1i,t} \), \( X_{2i,t} \), and \( X_{3i,t} \) denote the distances to agricultural area, forest, and river, respectively. Each neural network component follows the same architecture and is defined as
\[
ANN_{j,t}(X_{ji,t}) = W_2 \, \text{ReLU}(W_1 X_{ji,t} + b_1) + b_2,
\]
where \( W_1 \in \mathbb{R}^{3 \times 1} \) and \( b_1 \in \mathbb{R}^{3} \) correspond to the first linear layer, and \( W_2 \in \mathbb{R}^{8 \times 3} \) and \( b_2 \in \mathbb{R}^{8} \) correspond to the second linear layer. The first layer, implemented as \texttt{Linear(1,3)}, takes a single scalar distance value as input and projects it into a three-dimensional latent space. The subsequent ReLU activation introduces non-linearity, enabling the network to model asymmetric and threshold-like responses to distance. The second layer, \texttt{Linear(3,8)}, further transforms these features into an eight-dimensional representation that is passed to the structured additive predictor. In contrast to Model~1, where the neural components act as simple linear embeddings, the inclusion of a non-linear activation and an additional hidden layer allows Model~2 to capture more intricate relationships between distance variables and residential presence. At the same time, each neural network remains strictly associated with a single predictor, preventing interactions between distances and preserving a clear attribution of effects.

\item \textit{Model 3, unstructured deep regression with image and tabular input.}

The third specification extends the SDDR framework by incorporating local spatial context derived from image data in addition to the tabular distance-based predictors. While the structured component of the model remains unchanged and continues to describe the marginal effects of distances to agricultural area, forest, and river through smooth spline functions, this model introduces an unstructured deep learning component designed to extract information from neighborhood-level land cover configurations. Formally, the structured part of the predictor consists of smooth functions \( f_{1,t}(X_{1i,t}) \), \( f_{2,t}(X_{2i,t}) \), and \( f_{3,t}(X_{3i,t}) \), which retain their interpretability and describe how the probability of residential presence varies as a function of distance to each land-use category. These terms ensure continuity with the previous model specifications and allow for direct comparison of the estimated distance effects across models. The novelty of this specification lies in the neural network component, denoted \( ANN_{1,t}(U_{i,t}) \), which processes unstructured image-derived information. For each pixel \( i \) at time \( t \), a square neighborhood \( N_r(i) \) of radius \( r \) is extracted from the classified land cover map. Each pixel within this neighborhood is associated with a categorical label corresponding to one of the land cover classes, namely urban, forest, agriculture, river, or other. The resulting local patch \( U_{i,t} \) thus represents a spatial snapshot of the surrounding landscape configuration. The model is defined as
\begin{equation}
    \begin{aligned}
        g(E(Y_{i,t})) &=
        \log\left(\frac{p_{i,t}}{1 - p_{i,t}}\right) \\ &=
        \beta_0
        + f_{1,t}(X_{1i,t})
        + f_{2,t}(X_{2i,t})
        + f_{3,t}(X_{3i,t})
        + ANN_{1,t}(U_{i,t}),
    \end{aligned}
\end{equation}
where \( U_{i,t} \in \mathbb{R}^{p \times p} \) denotes the land cover labels within the neighborhood \( N_r(i) \), with \( p = 2r + 1 \) and cardinality \( p^2 \). The neural network component first flattens the neighborhood matrix into a vector, which is then mapped to a lower-dimensional latent representation through a linear transformation followed by a ReLU activation,
\[
ANN_{1,t}(U_{i,t}) = W_2 \, \text{ReLU}(W_1 \, \text{Flatten}(U_{i,t}) + b_1) + b_2,
\]
where \( W_1 \in \mathbb{R}^{a \times p^2} \) and \( b_1 \in \mathbb{R}^{a} \) define the first linear layer, and \( W_2 \) and \( b_2 \) map the extracted features to the additive predictor. In practice, the dimension \( a \) is chosen as a function of the neighborhood size to balance representational capacity and computational efficiency. This unstructured component allows the model to capture spatial arrangements and contextual patterns that are not explicitly encoded in distance variables, such as clustering of urban pixels, edge effects between land-use classes, or heterogeneous transition zones. Importantly, these image-based features enter the predictor additively, preserving the interpretability of the structured distance effects while enriching the model with local spatial context. By combining smooth distance-based effects with neighborhood-level image representations, this specification provides a unified framework for analyzing both global spatial gradients and local spatial structures. It therefore enables a more comprehensive representation of urbanization patterns, particularly in situations where proximity alone is insufficient to characterize land-use dynamics.
\end{itemize}

These three SDDR specifications illustrate the versatility of the framework in accommodating increasing levels of structural and data complexity. Model~1 can be viewed as a conservative extension of generalized additive models, in which smooth distance-based effects are complemented by shallow neural components that allow limited feature transformation while preserving a high degree of interpretability. Model~2 builds upon this structure by introducing deeper neural architectures for each distance predictor, enabling more flexible and non-linear representations of spatial effects without departing from a purely tabular data setting. Model~3 further extends the framework by integrating unstructured image-based information alongside the structured distance variables, allowing the model to explicitly capture local spatial configurations and neighborhood-level land cover patterns that cannot be represented by distance measures alone.

Beyond differences in model architecture, the SDDR framework also enables the incorporation of temporal context. In addition to models fitted using contemporaneous predictors only, we conducted experiments in which temporally lagged information was introduced. In this setting, predictors derived from the immediately preceding time step were included alongside current-year variables, both in the form of tabular distance measures and neighborhood-based image representations. This strategy allows the models to exploit short-term temporal dependencies in land use dynamics, reflecting the fact that urbanization processes are strongly conditioned by recent land cover states.

By comparing the predictive performance and inferred effects of these SDDR variants against traditional GLM and GAM baselines, we assess the extent to which progressively richer neural representations and the inclusion of temporal context improve model accuracy while maintaining interpretability. Together, these comparisons provide insight into the trade-offs between model complexity, explanatory power, and the ability to capture spatio-temporal patterns underlying land use change.

\subsection{Spatial cross-validation and evaluation metrics}\label{subsec:cv}

The robustness and accuracy of our models were evaluated using spatial cross-validation techniques, designed to account for spatial dependencies in the data and to validate model performance across different regions within the study area. To implement this, we divided the map into spatial blocks and performed cross-validation by partitioning the Donau-Ries study area into four distinct sections.

Spatial cross-validation requires careful handling of spatio-temporal dependencies to avoid artificially high performance estimates. Standard $k$-fold cross-validation, though commonly used, may yield overly optimistic accuracy by allowing spatially correlated observations to appear in both the training and test sets, thereby violating the assumption of independent data. This issue is particularly acute in spatial data, where adjacent regions often share underlying patterns due to spatial autocorrelation.

In our approach, we used block-based cross-validation, a strategy that excludes neighboring regions from the training set when validating a specific area, thus minimizing spatial autocorrelation biases. Such an approach aligns with the guidelines proposed by \cite{roberts2017cross}, who emphasize that spatially aware methods such as leave-location-out or spatial $k$-fold cross-validation help to prevent overfitting in models with spatial dependencies. A recent review by \cite{otto2024review} further underscores the importance of using specialized cross-validation for spatial and spatio-temporal data. This review details effective regularized estimation methods and cross-validation techniques tailored for spatial econometrics and geostatistics models, and it specifically recommends block-based cross-validation schemes in contexts where spatial autocorrelation is present.

To complement the spatial cross-validation framework, model performance was assessed using a set of standard classification metrics that provide complementary perspectives on predictive quality. These metrics are accuracy, precision, recall, and the F1-score. Accuracy measures the overall proportion of correctly classified pixels, offering a global view of model performance but potentially masking class imbalance effects. Precision quantifies the reliability of positive predictions by evaluating the proportion of correctly identified residential pixels among all pixels predicted as residential, making it particularly relevant for assessing over-prediction in urbanization modeling. Recall, also referred to as sensitivity, measures the ability of a model to correctly identify actual residential areas, thereby capturing how well urban expansion is detected across space. The F1-score, defined as the harmonic mean of precision and recall, provides a balanced summary of model performance by jointly penalizing false positives and false negatives. Using this combination of metrics ensures a robust and nuanced evaluation of the models, allowing performance to be interpreted not only in terms of overall accuracy but also with respect to the trade-offs between detection capability and prediction reliability, which is especially critical in spatially structured land-use data.

% =============================================================================
\section{Results}\label{sec3}

\subsection{Generalized linear model trends and coefficients}

We applied GLM, GAM and SDDR analyses over four years, namely 2023, 2015, 2009 and 2004, to identify factors associated with the likelihood of settlement areas with various geographical characteristics. The results reveal evolving patterns in the association between settlement distributions and distances to agricultural areas, rivers, and forests.

In the GLM framework with a logit link, the estimated coefficients quantify the effect of each explanatory variable on the log-odds of urbanization. To facilitate interpretation, these coefficients are often expressed as odds ratios obtained by exponentiating the estimated parameters. An odds ratio greater than 1 indicates that an increase in the predictor is associated with higher odds of urbanization, whereas a value below 1 implies a decrease in the odds. In this context, odds ratios provide a direct way to assess how changes in distance to landscape features relate to the relative likelihood of urban development.

The coefficients associated with distance to agricultural areas are positive for all years considered, with values ranging from 1.11 in 2009 to 1.44 in 2015. This indicates that the probability of urbanization increases as the distance from agricultural land increases. In terms of odds ratios, these coefficients correspond to values between approximately 3.04 and 4.23. For example, the coefficient of 1.33 estimated for 2004 corresponds to an odds ratio of about 3.78, meaning that a one-unit increase in the log-distance to agricultural areas multiplies the odds of urbanization by nearly four. This strong and persistent association suggests that settlements tend to develop away from agricultural land, possibly reflecting land-use competition or regulatory constraints limiting urban expansion on productive agricultural areas.

Distance to forested areas also exhibits positive coefficients across all years, ranging from 1.15 in 2009 to 1.40 in 2015. These coefficients correspond to odds ratios ranging from approximately 3.17 to 4.05, indicating that urbanization becomes more likely as the distance from forested areas increases. This pattern suggests that settlements tend to concentrate in more open landscapes, while forested zones may act as physical or regulatory constraints on urban expansion. The temporal stability of this relationship indicates a consistent association between forest proximity and settlement location throughout the study period.

The effect of distance to rivers varies slightly across the considered years. In 2004 and 2015, the estimated coefficients are positive but relatively small, with values of 0.089 and 0.049, respectively, corresponding to odds ratios of approximately 1.09 and 1.05. These values indicate that increasing the distance from rivers slightly increases the odds of urbanization during these periods. In contrast, the coefficients for 2009 and 2023 are negative, with values of $-0.131$ and $-0.092$, corresponding to odds ratios of approximately 0.88 and 0.91. These values indicate that increasing the distance from rivers decreases the odds of urbanization in these years, suggesting a greater tendency for settlements to occur closer to rivers. Although the magnitude of these effects remains relatively modest compared with those associated with agricultural or forest distances, the results highlight the potentially complex role of river proximity in settlement patterns, which may reflect both environmental constraints such as flood risk and the attractiveness of riverine landscapes for residential development.

\begin{table}
  \centering
  \caption{Results of the generalized linear model fitted to binary data on the spatial location of settlements for the years 2004, 2009, 2015 and 2023, with explanatory variables transformed into log-distance to landscape features (agriculture, forest, river).}
  \label{tab1}
  \footnotesize
  \begin{tabular}{@{}l@{\extracolsep{6pt}}rrrr@{}}
    \hline
     &  Estimate & Std. error & $z$-value & $p$-value \\
    \multicolumn{5}{l}{\emph{\textcolor{darkyellow}{Estimated GLM 2004}}}  \\
    \\
    (Intercept) & $-12.22622$  & 0.19674  & $-62.145$   & $<2 \times 10^{-16}$ \\
    Distance to agriculture (log-m) &  1.33478 & 0.01861  & 71.715  & $<2 \times 10^{-16}$ \\
    Distance to forest (log-m) & 1.34021 & 0.02395   & 55.953  & $<2 \times 10^{-16}$ \\
    Distance to river (log-m) & 0.08898 & 0.01527    & 5.826  & $<5.68 \times 10^{-09}$\\
    \\
    \multicolumn{5}{l}{\emph{\textcolor{green}{Estimated GLM 2009}}}  \\
    \\
    (Intercept) & $-9.76060$ & 0.17125  & $-56.996$    & $<2 \times 10^{-16}$ \\
    Distance to agriculture (log-m) & 1.11125 & 0.01630  & 68.181 & $<2 \times 10^{-16}$ \\
    Distance to forest (log-m) & 1.15481    & 0.02261 & 51.065 & $<2 \times 10^{-16}$ \\
    Distance to river (log-m) & $-0.13093$ & 0.01467  & $-8.923$   &$ <2 \times 10^{-16}$  \\
    \\
    \multicolumn{5}{l}{\emph{\textcolor{cyan}{Estimated GLM 2015}}}  \\
    \\
    (Intercept) & $-12.74058$  & 0.20109  & $-63.356$ & $<2 \times 10^{-16}$ \\
    Distance to agriculture (log-m) & 1.44145  & 0.02113  &68.209  & $<2 \times 10^{-16}$ \\
    Distance to forest (log-m) & 1.39524 & 0.02468   & 56.544  & $<2 \times 10^{-16}$ \\
    Distance to river (log-m) & 0.04858 & 0.01578 & 3.079 & 0.00207 \\
    \\
    \multicolumn{5}{l}{\emph{\textcolor{blue}{Estimated GLM 2023}}}  \\
    \\
    (Intercept) & $-11.16013$  & 0.19506  & $-57.215$   & $<2 \times 10^{-16}$ \\
    Distance to agriculture (log-m) &  1.13067 & 0.01618  & 69.863  & $<2 \times 10^{-16}$ \\
    Distance to forest (log-m) &1.36857     & 0.02812  & 48.674   & $<2 \times 10^{-16}$ \\
    Distance to river (log-m) & $-0.09213$    & 0.01361   & $-6.769$  & $1.3 \times 10^{-11}$ \\
    \hline
    \multicolumn{5}{l}{\emph{Goodness-of-fit}}  \\
    Adj. $R^2$ (in-sample) & 0.6111 & & &  \\
    Accuracy (cross-validation)  & 0.7556 & & & \\
    Precision (cross-validation) & 0.3784 & & & \\
    Recall (cross-validation) & 0.9774 & & & \\
    F1 score (cross-validation) & 0.5200 & & & \\
    \hline
  \end{tabular}
\end{table}

As shown in Table~\ref{tab1}, the GLM was fitted separately for four key years to predict the probability of urban pixels based on log-transformed distances to major landscape features, namely agricultural areas, forests and rivers. The results show that the distances to agricultural areas and forests exert a strongly positive and significant influence on the probability of settlement for all years. The effect of distance from the river, by contrast, varies from year to year. It is weakly positive in 2004 and 2015, but becomes significantly negative in 2009 and 2023, which suggests an evolving pattern in the attractiveness of areas close to rivers. The significant negative intercepts indicate an overall low probability of settlement in the absence of proximity to these features. The model shows a moderate in-sample fit with an adjusted $R^2$ of 0.61. It nevertheless demonstrates a good generalization capacity on the cross-validation data, with an accuracy of 0.75 and a particularly high recall of 0.98, indicating that the model is very effective at detecting settlement presence. However, the relatively low precision of 0.38 suggests that a substantial number of predicted presences correspond to false positives. This imbalance between recall and precision is also reflected in the moderate F1 score of 0.52.

\subsection{Generalized additive model analysis of smooth terms for distances to agricultural area, river and forest}

\begin{figure}[h!]
\centering
\subfigure[Distance to agricultural area.]{\includegraphics[width=0.31\textwidth]{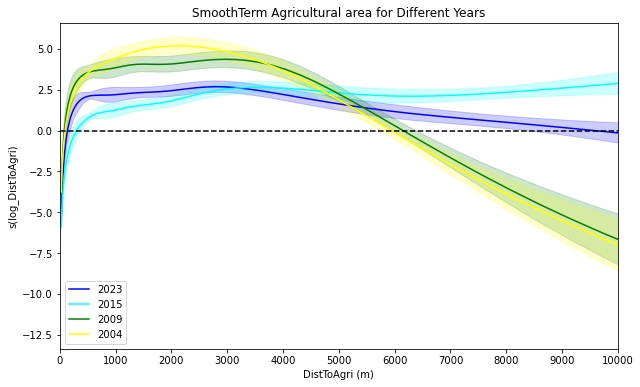}}
\hfill
\subfigure[Distance to forest.]{\includegraphics[width=0.31\textwidth]{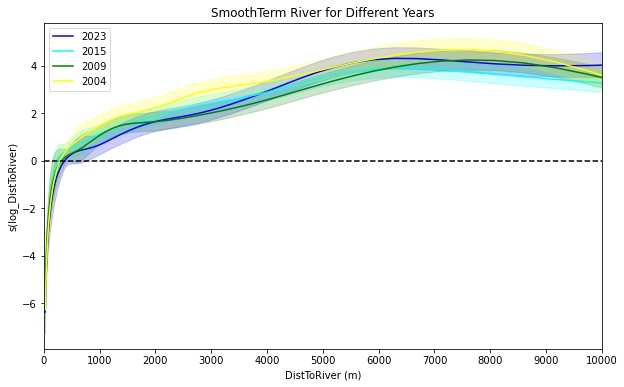}}
\hfill
\subfigure[Distance to river.]{\includegraphics[width=0.31\textwidth]{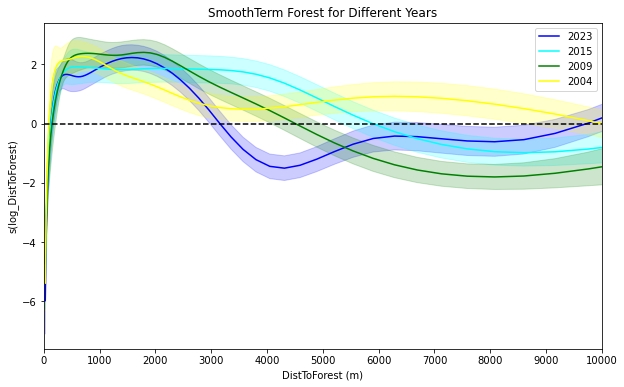}}
\caption{Temporal analysis of the estimated GAM smooth terms for different landscape features across the years 2004, 2009, 2015, and 2023.}
\label{Label3}
\end{figure}

\begin{table}
  \centering
  \caption{Results of the generalized additive model for the years 2004, 2009, 2015 and 2023, including smoothed effects of log distances to landscape features (agriculture, forest, river) and spatial coordinates, as well as effective degrees of freedom and significance statistics.}
  \label{tab02}
  \footnotesize
  \begin{tabular}{@{}l@{\extracolsep{6pt}}rrrr@{}}
     \hline
     Variables & edf & Ref.df & Chi.sq & $p$-value \\
    \multicolumn{5}{l}{\emph{\textcolor{darkyellow}{Estimated GAM 2004}}}  \\
    \\
    s(Distance to agriculture (log-m)) &  7.789  & 7.968 & 1161.1  & $<2 \times 10^{-16}$ \\
    s(Distance to forest (log-m)) & 8.052  & 8.577  & 401.5 & $<2 \times 10^{-16}$ \\
    s(Distance to river (log-m)) & 8.993 & 9.000 & 448.0   & $<2 \times 10^{-16}$ \\
     s(x\_coordinate, y\_coordinate) & 28.833  & 28.995  & 641.5  & $<2 \times 10^{-16}$ \\
     \\
    \multicolumn{5}{l}{\emph{\textcolor{green}{Estimated GAM 2009}}}  \\
    \\
    s(Distance to agriculture (log-m)) & 6.764  & 7.007 &1161.3  & $<2 \times 10^{-16}$ \\
    s(Distance to forest (log-m)) &8.810  & 8.977 & 509.8  & $<2 \times 10^{-16}$ \\
    s(Distance to river (log-m)) & 8.923   & 8.995 & 398.2  & $<2 \times 10^{-16}$  \\
    s(x\_coordinate, y\_coordinate) & 28.907  & 28.998 & 892.3   & $<2 \times 10^{-16}$ \\
    \\
    \multicolumn{5}{l}{\emph{\textcolor{cyan}{Estimated GAM 2015}}}  \\
    \\
    s(Distance to agriculture (log-m)) & 3.281  & 3.842   & 1332.0   & $<2 \times 10^{-16}$ \\
    s(Distance to forest (log-m)) &8.080   & 8.593  & 519.4  & $<2 \times 10^{-16}$ \\
    s(Distance to river (log-m)) & 8.890  & 8.992  & 330.5  & $<2 \times 10^{-16}$\\
     s(x\_coordinate, y\_coordinate) & 28.968 & 29.000  & 872.2   & $<2 \times 10^{-16}$ \\
     \\
    \multicolumn{5}{l}{\emph{\textcolor{blue}{Estimated GAM 2023}}}  \\
    \\
    s(Distance to agriculture (log-m)) &  4.371  & 5.111  & 1274.3  & $<2 \times 10^{-16}$ \\
    s(Distance to forest (log-m)) &8.499  & 8.853   & 381.2  & $<2 \times 10^{-16}$ \\
    s(Distance to river (log-m)) & 8.895   & 8.994   & 443.7   & $<2 \times 10^{-16}$ \\
     s(x\_coordinate, y\_coordinate) & 28.999  & 29.000  & 997.1   & $<2 \times 10^{-16}$ \\
    \hline
    \multicolumn{5}{l}{\emph{Goodness-of-fit}}  \\
    Adj. $R^2$ (in-sample) & 0.9297 &  & & \\
    Accuracy (cross-validation)  & 0.6979 & & & \\
    Precision (cross-validation) & 0.3543 & & & \\
    Recall (cross-validation) & 0.6240 & & & \\
    F1 score (cross-validation) &  0.3327 & & & \\
    \hline
  \end{tabular}
\end{table}

The temporal analysis of smooth terms within the GAM examines how the estimated non-linear effects of landscape variables on residential development evolve. Each smooth term represents a flexible, data-driven function of a distance variable, allowing the relationship between distance and urbanization probability to vary across space and across years. By comparing these smooth functions for different time periods, this analysis reveals shifts in settlement patterns from 2004 to 2023 (see Figure~\ref{Label3}).

Because the GAM is fitted using a logistic link function, the smooth terms are expressed on the logit scale. The logit effect corresponds to the additive contribution of a predictor to the log-odds of urbanization. Positive values indicate an increased probability of urban development, while negative values indicate a reduced likelihood. A logit value of zero corresponds to no effect relative to the baseline, whereas large positive or negative values indicate strong attraction or repulsion effects, respectively.

The smooth term associated with distance from agricultural areas highlights a strong attractiveness of locations in close proximity to agricultural areas, particularly in 2004 and 2009. For these years, the logit effect reaches values greater than 5 between 1,000 and 2,000 m, corresponding to a very high probability of urbanization in these zones. Beyond 2,000 to 3,000 m, this effect decreases markedly, and for the earlier years the trend reverses after 4,000 m, with negative logit values, indicating a reduced probability of urbanization at greater distances. In contrast, the curves for 2015 and 2023 show a general attenuation of this effect, with the smooth term stabilizing around zero beyond 3,000 m. This temporal evolution suggests a reduced dependence on proximity to agricultural infrastructure in more recent periods, possibly reflecting changes in mobility, land availability, or a broader diversification of the factors behind urbanization.

The smooth term related to distance from forests exhibits a strongly increasing pattern across all years. At very short distances below 500 m, the effect is markedly negative, especially in 2004 where the logit reaches approximately $-6$, corresponding to an odds ratio below 0.002. This indicates a strong avoidance of immediate forest proximity. As distance increases, the negative effect weakens and becomes positive around 1,500 m, reaching peak values between 6,000 and 8,000 m, with logits between 2 and 4, particularly in 2004 and 2009. This profile reflects a preference for open or semi-open areas where forests remain accessible as landscape or resource features, without the constraints associated with direct adjacency such as shading, humidity, or land-use restrictions. In 2015 and 2023, the smoother rise of the curves suggests a diminished sensitivity to forest proximity in more recent periods.

The smooth term associated with distance from rivers displays a complex, non-monotonic relationship. At very short distances below 500 m, the effect is generally negative, with logit values below $-2$ in 2009 and 2023, indicating an avoidance of flood-prone or risk-exposed areas. Between approximately 500 and 3,000 m, the effect becomes positive, reaching maximum logit values between 2 and 3 depending on the year, which reflects an optimal balance between proximity to water-related amenities and exposure to environmental risks. At larger distances above 5,000 m, the temporal patterns diverge. In 2004, the effect remains neutral or slightly positive, whereas in 2009 and 2023 it becomes negative again. This suggests a more recent decline in the attractiveness of areas far from watercourses, potentially linked to increased valuation of water access, recreational amenities, or landscape quality in contemporary urban development processes.

The GAM results reveal that the responses of urbanization patterns to the spatial environment are far from linear. The attractiveness of agricultural areas decreases with distance, but also over time. Forests continue to act as an immediate repellent, but this repulsion becomes an asset over longer distances. Finally, rivers illustrate an ambivalent relationship, where attractiveness is conditioned by a balance between access and risk. Changes in the curves between 2004 and 2023 show a trend towards greater complexity in the spatial logic of location, marked by less dependence on single factors such as the agricultural area and finer management of natural constraints such as forest and river. This suggests a gradual refinement of development choices, most probably linked to more controlled and planned urbanization.

The GAM was applied with the same explanatory variables as the GLM, but allowing for non-linear effects via smoothed terms, and adding a smooth two-dimensional term based on spatial coordinates. The effective degrees of freedom associated with the smoothed terms are all well above 1, confirming the presence of non-linear relationships between distances to landscape features and the probability of settlement. Distance to agricultural area, in particular, shows structures of varying complexity depending on the year, with effective degrees of freedom ranging from 3.281 in 2015 to 7.789 in 2004, indicating strong spatial and temporal variation in the agricultural effect. The smooth two-dimensional term is also very complex, with almost 29 degrees of freedom for all years (see Table~\ref{tab02}), highlighting a strong spatial structuring of the settlements that is not explained by the environmental variables alone. All terms are highly significant ($p<2 \times 10^{-16}$). The GAM shows a very good in-sample fit, with an average adjusted $R^2$ of 0.93. However, its performance in cross-validation is moderate, with an accuracy of 0.70, a recall of 0.62, a precision of 0.35 and an F1 score of 0.33.

The comparison of GLM and GAM reveals a classic tension between goodness of fit and generalizability. The GAM clearly outperforms the GLM in terms of in-sample goodness of fit, with a mean adjusted $R^2$ of 0.93 versus 0.61, thanks to its ability to model complex non-linear relationships and explicit spatial effects. However, this increased flexibility seems to compromise its performance in cross-validation, with lower scores than the GLM on all indicators, particularly recall (0.62 versus 0.98) and F1 score (0.33 versus 0.52). The GLM, although more rigid, seems to generalize better, particularly when it comes to identifying the areas actually occupied by settlement. This contrast suggests that, despite the appeal of the GAM for exploring complex structures, the GLM offers a more robust compromise in predictive contexts, particularly where the ability to avoid false positives is essential. In summary, the GAM is more informative about underlying spatial structure, while the GLM is more reliable for operational predictions.

\subsection{Partial effects for distances to agricultural area, forest and river in the SDDR framework}

\begin{figure}[h!]
\centering
\subfigure[Distance to agricultural area.]{\includegraphics[width=0.31\textwidth]{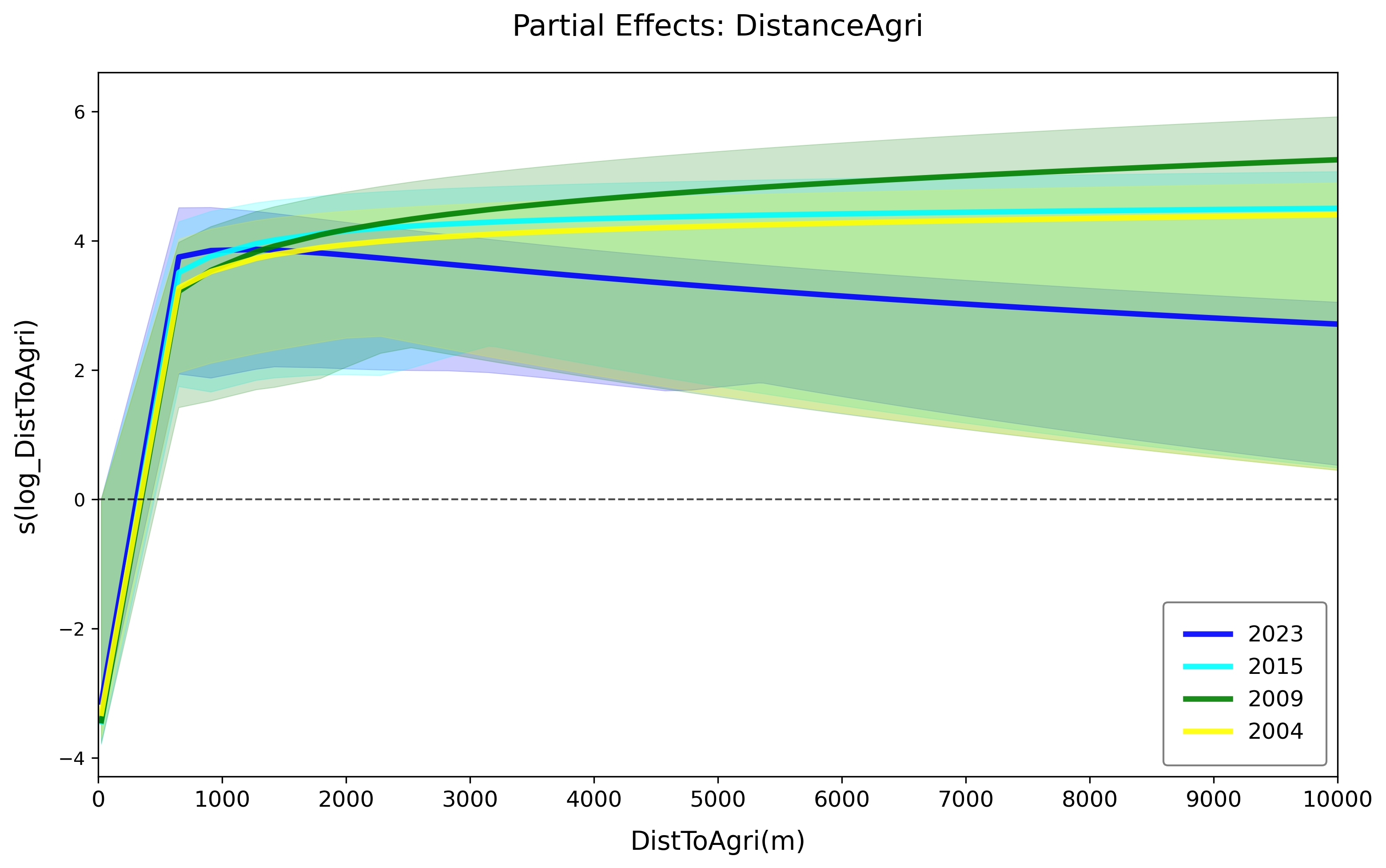}}
\hfill
\subfigure[Distance to forest.]{\includegraphics[width=0.31\textwidth]{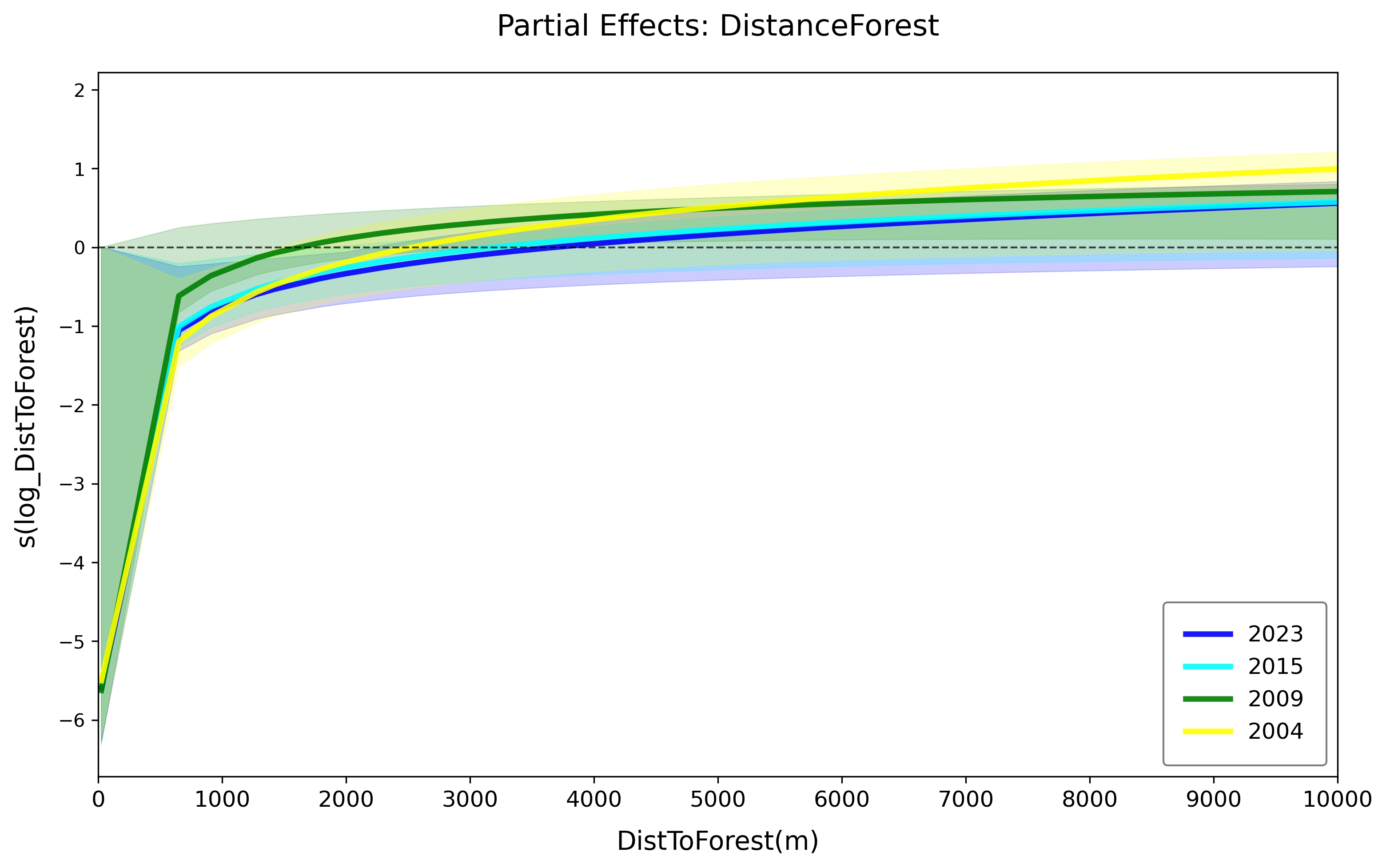}}
\hfill
\subfigure[Distance to river.]{\includegraphics[width=0.31\textwidth]{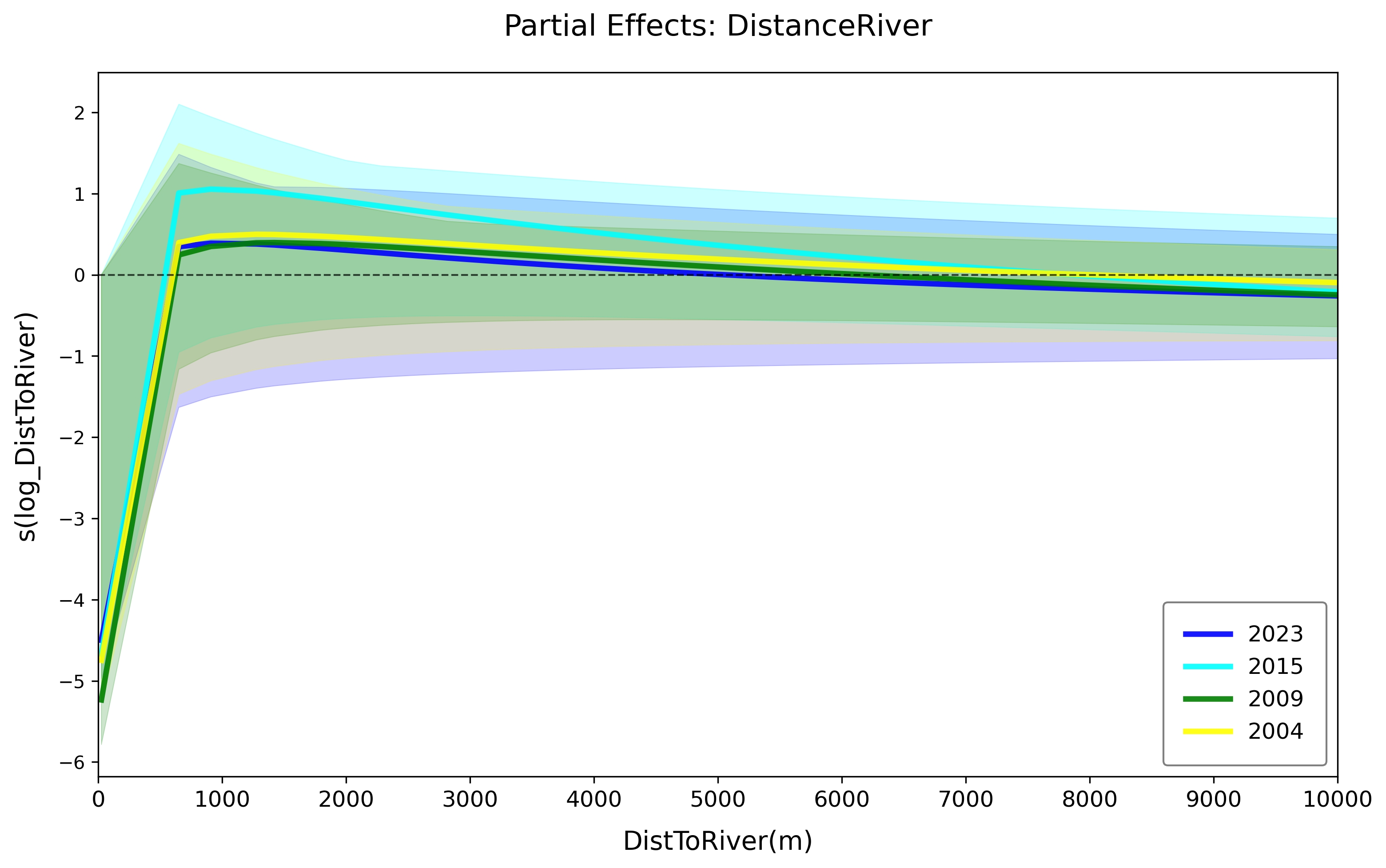}}
\caption{Partial effects from SDDR Model~3 with $17\times17$ pixel windows for different landscape features across the years 2004, 2009, 2015, and 2023. The $x$-axis shows the various distances in meters, while the $y$-axis shows the estimated partial effects, that is, the relative contribution of each variable to the logit of the probability of urbanization.}
\label{Label4}
\end{figure}

The partial effects derived from the SDDR model (see Figure~\ref{Label4}) represent the isolated contribution of each predictor to the linear predictor on the logit scale, while holding all other variables constant. In other words, a partial effect curve describes how variations in a single distance variable, such as distance to agricultural areas, forests, or rivers, modify the log-odds of urbanization, after accounting for the remaining structured and deep components of the model. Positive values of the partial effect indicate an increased probability of urban development, whereas negative values correspond to a reduced likelihood. By examining these curves across multiple years, the SDDR framework enables a direct assessment of how the estimated influence of landscape features on settlement patterns evolves over time.

The partial effects associated with distance to agricultural areas reveal a consistently strong and positive association between proximity and urbanization. Across all years, the probability of settlement increases sharply within the first kilometer, confirming the high attractiveness of areas close to agricultural land. This effect intensifies over time, reaching its highest magnitude in 2023, which suggests a growing functional role of agricultural areas in recent urban dynamics. Such reinforcement may reflect improved accessibility, the availability of open land, or the increasing value of peri-urban agricultural landscapes providing ecosystem services or visual amenities. While earlier years exhibit a similar overall pattern, the increasing amplitude of the curves indicates that agricultural proximity has become a more decisive correlate of urban expansion.

In contrast, the partial effects for distance to forests display a markedly different structure. Locations situated very close to forested areas are systematically associated with negative logit values, highlighting a reduced likelihood of urban presence. This repulsive effect is strongest at short distances and can be interpreted as a response to physical constraints, limited accessibility, shading, or land-use regulations. Beyond approximately 4,000 meters, however, the effect becomes positive, suggesting a preference for more open environments or transitional zones where forests remain accessible without imposing direct constraints. Importantly, this pattern remains remarkably stable across all study years, indicating that the role of forests in shaping settlement choices has been both persistent and spatially consistent over time.

The influence of distance to rivers exhibits a more complex and non-monotonic behavior. At very short distances below 500 meters, the partial effects are predominantly negative, reflecting a clear avoidance of areas exposed to hydrological risks such as flooding. At intermediate distances, roughly between 500 and 2,000 meters, the effect turns positive, identifying a zone of optimal attractiveness where the benefits of proximity to watercourses, such as landscape value, microclimate, and accessibility, are maximized without incurring significant risk. This favorable window is particularly pronounced before 2015. In more recent years, the magnitude of this positive effect decreases slightly, suggesting either improved flood-risk awareness, stricter planning regulations, or changing preferences in urban development strategies.

Overall, the partial effects obtained from the SDDR model provide a nuanced and temporally explicit view of how landscape features relate to urbanization patterns. By isolating the contribution of each predictor while retaining the flexibility of deep learning components, the model captures both stable structural regularities and evolving spatial dynamics. While pointwise confidence intervals tend to overlap across years, the overall shapes and inflexions of the effect functions remain robust, allowing for meaningful interpretation of the underlying spatial preferences. Together, these results highlight both enduring environmental constraints and evolving land-use strategies, reflecting a trajectory towards more informed, risk-aware, and infrastructure-oriented urban development in the region.

\subsection{Impact of temporal images on partial effects and predictive performance}

\begin{table}[h!]
\centering
\caption{Classification metrics for SDDR models with temporal context across different image window sizes.}
\label{Tab:TemporalWindows}
\small
\begin{tabular}{|c|c|c|c|c|c|}
\hline
\textbf{Window} & \textbf{Year} & \textbf{Accuracy} & \textbf{Precision} & \textbf{Recall} & \textbf{F1-score} \\
\hline
\multirow{6}{*}{$5\times5$}
& 2009 & 0.9164 & 0.6821 & 0.5353 & 0.5895 \\
& +2004 & \textbf{0.9593} & \textbf{0.8037} & \textbf{0.7529} & \textbf{0.7748} \\\hhline{|~|-|-|-|-|-|}
& 2015 & \textbf{0.9489} & \textbf{0.8375} & \textbf{0.7643} & \textbf{0.7991} \\
& +2009 & 0.9438 & 0.8026 & 0.6965 & 0.7441 \\\hhline{|~|-|-|-|-|-|}
& 2023 & 0.9223 & \textbf{0.7811} & 0.6844 & \textbf{0.7014} \\
& +2015 & \textbf{0.9378} & 0.7635 & \textbf{0.7245} & 0.6980 \\
\hline
\multirow{6}{*}{$17\times17$}
& 2009 & 0.9483 & 0.7372 & \textbf{0.6857} & \textbf{0.7082} \\
& +2004 & \textbf{0.9557} & \textbf{0.7754} & 0.6514 & 0.7077 \\\hhline{|~|-|-|-|-|-|}
& 2015 & \textbf{0.9490} & \textbf{0.7940} & \textbf{0.7125} & \textbf{0.7501} \\
& +2009 & 0.9328 & 0.7207 & 0.6070 & 0.6585 \\\hhline{|~|-|-|-|-|-|}
& 2023 & 0.9415 & 0.7625 & \textbf{0.7106} & 0.7058 \\
& +2015 & \textbf{0.9484} & \textbf{0.8068} & 0.6851 & \textbf{0.7302} \\
\hline
\multirow{6}{*}{$33\times33$}
& 2009 & 0.9133 & 0.6592 & 0.4968 & 0.5520 \\
& +2004 & \textbf{0.9244} & \textbf{0.6681} & \textbf{0.5679} & \textbf{0.5994} \\\hhline{|~|-|-|-|-|-|}
& 2015 & \textbf{0.9251} & 0.6753 & \textbf{0.5381} & \textbf{0.5959} \\
& +2009 & 0.9154 & \textbf{0.6761} & 0.5178 & 0.5834 \\\hhline{|~|-|-|-|-|-|}
& 2023 & \textbf{0.9385} & 0.6973 & \textbf{0.5997} & \textbf{0.6272} \\
& +2015 & 0.9186 & \textbf{0.7371} & 0.5400 & 0.5590 \\
\hline
\end{tabular}
\end{table}

To assess the contribution of temporal context in the SDDR framework, we analyzed both the evolution of partial effects and the predictive performance of models trained with and without historical imagery. For each target year, namely 2009, 2015, and 2023, two model configurations were compared. The first is a baseline model relying solely on data from the focal year, and the second is an augmented model incorporating contextual imagery from the immediately preceding period, respectively 2004, 2009, and 2015. This design allows us to jointly evaluate how temporal information influences the estimated relationships between landscape features and urbanization and overall model accuracy.

\begin{figure}[h!]
\centering
\includegraphics[scale=0.25]{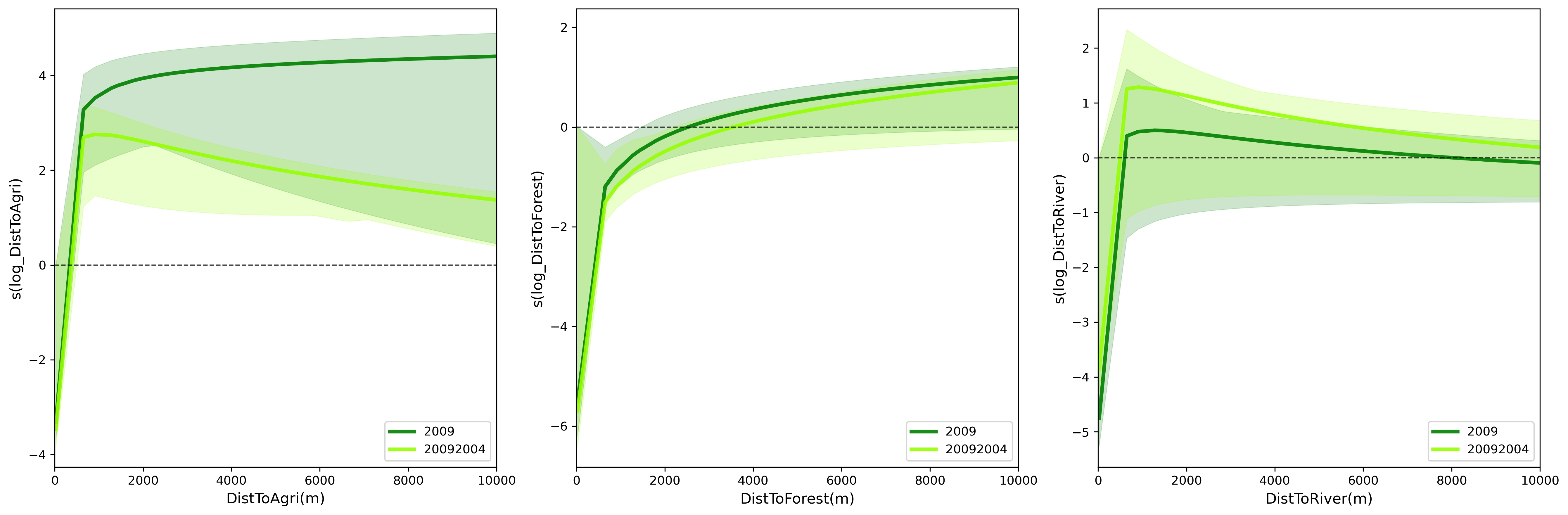}
\caption{Influence of 2004 imagery on the partial effects for the year 2009. Comparison of smooth effects for 2009 alone, shown by the green curve, versus 2009 enriched with 2004 imagery, shown by the light green curve, highlighting the influence of past landscape context on the spatial relationships with agricultural area, forest, and river.}
\label{fig:temp2009}
\end{figure}

From a structural perspective, the inclusion of historical imagery leads to noticeably smoother and more stable partial effects. In the case of 2009 (Figure~\ref{fig:temp2009}), integrating the 2004 image results in more conservative distance effects compared to the model trained on 2009 data alone. For distances to agricultural areas, the augmented model exhibits a dampened increase in urbanization probability, suggesting that temporal context helps stabilize the estimated effect by mitigating short-term land cover variability. Similar regularization effects are observed for distances to forests and rivers, where the smooth terms become flatter and less erratic when historical imagery is included. Overall, these patterns indicate that temporal imagery acts as a form of contextual constraint, reducing noise and enhancing the robustness of the learned spatial relationships.

These structural improvements are largely reflected in the predictive performance of the models. When accounting for temporal dependence by including imagery from the previous period, performance metrics generally improve, particularly for smaller spatial contexts. As shown in Table~\ref{Tab:TemporalWindows}, for a $5 \times 5$ pixel image window, corresponding to $125 \times 125$ m, the F1-score for 2009 increases substantially from 0.59 to 0.77 when the 2004 image is added. In contrast, performance for 2015 slightly decreases when temporal imagery is included, likely reflecting temporal inconsistencies or transitional urban development dynamics that are not fully captured by information from the preceding period.

For larger spatial windows, for example $33 \times 33$ pixels in Table~\ref{Tab:TemporalWindows}, the benefits of temporal augmentation become less consistent and may even reverse. This suggests that overly broad spatial contexts can dilute locally relevant temporal signals, thereby reducing the added value of historical imagery. Taken together, these results indicate that the joint use of spatial and temporal information is most effective when applied at local scales, where historical context meaningfully complements contemporaneous landscape features.

\begin{figure}[h!]
\centering
\includegraphics[scale=0.25]{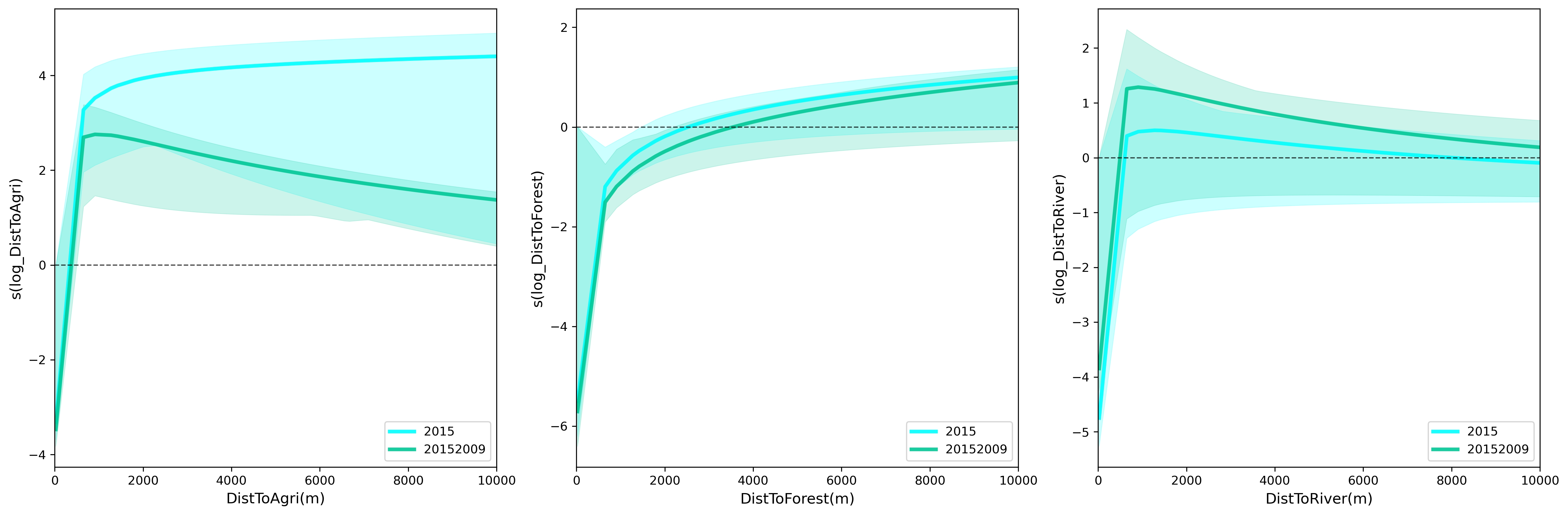}
\caption{Influence of 2009 imagery on the partial effects for the year 2015. Comparison of smooth effects for 2015 alone, shown by the cyan curve, versus 2015 enriched with 2009 imagery, shown by the turquoise green curve, highlighting the influence of past landscape context on the spatial relationships with agricultural area, forest, and river.}
\label{fig:temp2015}
\end{figure}

\begin{figure}[h!]
\centering
\includegraphics[scale=0.25]{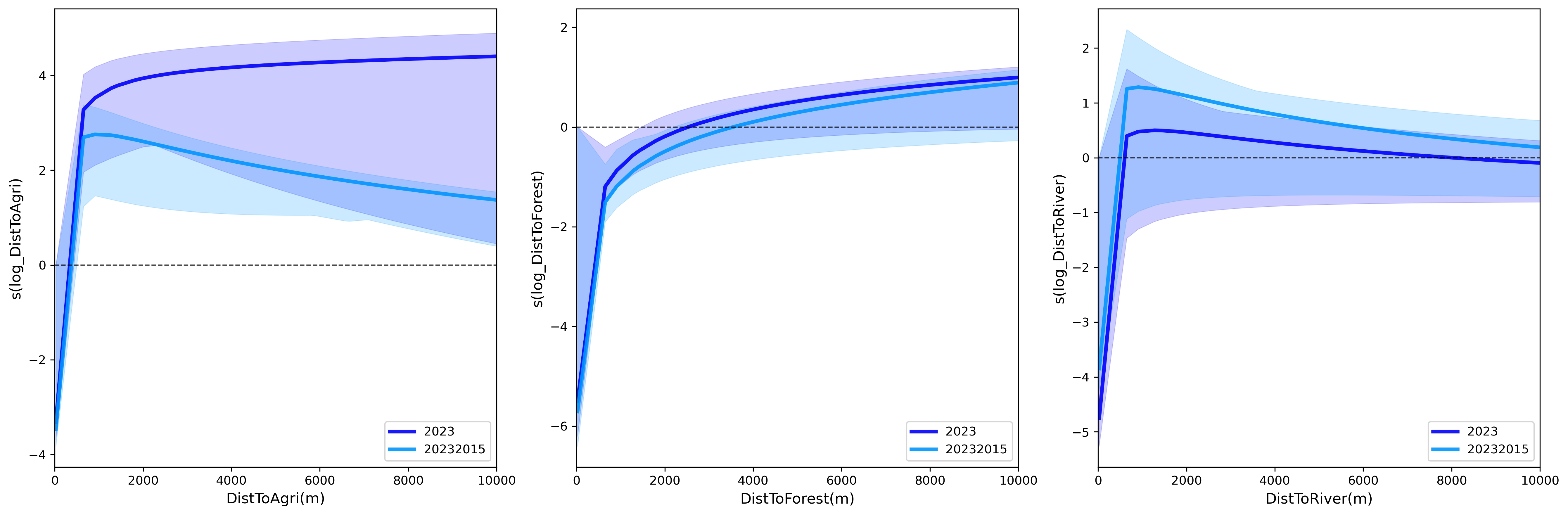}
\caption{Influence of 2015 imagery on the partial effects for the year 2023. Comparison of smooth effects for 2023 alone, shown by the blue curve, versus 2023 enriched with 2015 imagery, shown by the turquoise blue curve, highlighting the influence of past landscape context on the spatial relationships with agricultural area, forest, and river.}
\label{fig:temp2023}
\end{figure}

\subsection{Comparative analysis of all models}

Table~\ref{table} presents a comparative analysis of the different modeling approaches across four evaluation metrics, namely accuracy, precision, F1-score, and recall, aggregated over all studied years. The GLM outperforms the GAM in terms of accuracy (0.75 versus 0.70), recall (0.98 versus 0.62), and F1-score (0.52 versus 0.33), while both models exhibit similarly low precision (0.38 and 0.35), which suggests a high rate of false positives. Although the GLM performs better overall, the low precision in both cases points to a limited ability of these traditional models to finely discriminate urbanization patterns. The first versions of the SDDR model, Model~1 and Model~2, which are not yet enriched with image context, demonstrate poor overall performance with very low precision (0.18 to 0.19) and F1-scores (0.27 to 0.30), despite high recall values (0.98 to 0.96), indicating over-prediction of positive cases. In contrast, Model~3, which incorporates spatial image context, shows a substantial leap in performance. The smallest context window of $5 \times 5$ pixels yields high accuracy (0.93) and a well-balanced performance across all metrics, with precision of 0.77, F1-score of 0.70, and recall of 0.67. The best overall performance is achieved by the $17 \times 17$ pixel version of Model~3, which combines high accuracy (0.95), solid precision (0.75), and balanced F1-score (0.71) and recall (0.70), outperforming both traditional models and earlier SDDR versions. We note, however, that performance decreases slightly with an image size of $33 \times 33$ pixels. These findings underscore the added value of integrating moderate-scale spatial context into deep models and confirm that SDDR, especially when paired with image-based spatial features, offers a powerful and scalable alternative to more interpretable but less flexible classical statistical models.

\begin{table}[h!]
    \centering
    \caption{Performance comparison from cross-validation of models averaged across all years (2004, 2009, 2015, 2023).}
    \label{table}
    \begin{tabular}{|l|c|c|c|c|}
    \hline
     Model & Accuracy & Precision & F1-score & Recall \\ \hline
     GLM & 0.75 & 0.38 & 0.52 & \textbf{0.98} \\ \hline
     GAM & 0.70 & 0.35 & 0.33 & 0.62 \\ \hline
     SDDR Model 1 & 0.44 & 0.18 & 0.27 & \textbf{0.98} \\ \hline
     SDDR Model 2 & 0.48 & 0.19 & 0.30 & 0.96 \\ \hline
     SDDR Model 3, $5\times 5$ pixels & 0.93 & \textbf{0.77} & 0.70 & 0.67 \\ \hline
     SDDR Model 3, $17\times 17$ pixels & \textbf{0.95} & 0.75 & \textbf{0.71} & 0.70 \\ \hline
     SDDR Model 3, $33\times 33$ pixels & 0.92 & 0.67 & 0.60 & 0.56 \\ \hline
    \end{tabular}
\end{table}

% =============================================================================
\section{Discussion}\label{sec4}

The comparative analysis reveals several important methodological insights. First, while the GLM and the GAM provide stable and interpretable estimates, their limited precision indicates difficulties in distinguishing subtle spatial urbanization patterns. This suggests that distance-based environmental predictors alone may not fully capture the spatial complexity underlying residential expansion.

Second, the poor performance of the initial SDDR specifications highlights that increased model flexibility does not automatically translate into improved predictive accuracy. Without contextual spatial information, the deep component tends to over-predict positive cases, as reflected by high recall but low precision. This emphasizes the importance of meaningful feature construction in hybrid modeling frameworks.

The substantial improvement observed when incorporating image-based spatial context indicates that local neighborhood structure contains valuable information for modeling urban development patterns. Although certain window sizes achieve slightly better performance, supplementary analyses do not provide clear evidence of a systematic relationship between neighborhood extent and predictive accuracy. This highlights the importance of contextual spatial features while suggesting that model performance is not determined by window size alone.

From a statistical perspective, these findings illustrate the trade-off between interpretability and predictive flexibility. While classical structured models remain valuable for inference and stable effect estimation, semi-structured deep models provide tangible gains when spatial dependencies are complex and multi-scale.

A further point concerns the covariate set itself. The three landscape distances used throughout this article were selected because they are consistently available across the four reference years and because they remain directly interpretable in every model considered. They do not, however, exhaust the mechanisms that generate urban growth. Population dynamics, planning regulation, road accessibility, and land-market pressure all shape settlement location, and none of them enters the models estimated here. The comparison presented in this article should therefore be read as a methodological assessment carried out under a fixed and deliberately parsimonious covariate set, rather than as an attempt to identify the determinants of urban expansion in a causal sense. The gain obtained when image context is added supports this reading, because that gain measures precisely how much spatial information the three distances fail to capture on their own.

A second limitation concerns the geographical scope of the analysis. The study relies on a single region and on four reference years, which is sufficient for a controlled comparison of modeling frameworks but not for establishing the transferability of the estimated effects. Replicating the same protocol on additional regions with contrasting settlement histories would clarify which parts of the reported behavior are specific to Donau-Ries and which parts reflect properties of the modeling frameworks themselves.

% =============================================================================
\section{Conclusion}\label{sec5}

This study presents a comparative case study of urbanization patterns in Donau-Ries county, combining classical statistical models with semi-structured deep learning approaches applied to historical land cover maps. By jointly analyzing generalized linear models, generalized additive models, and semi-structured deep distributional regression, we were able to distinguish interpretable spatial effects while progressively integrating higher levels of spatial and temporal complexity.

Across all modeling frameworks, consistent spatial patterns emerge. Residential areas exhibit a strong attraction to agricultural areas, particularly at short distances, an effect that has intensified over time. In contrast, proximity to forests is associated with a persistent negative influence on settlement presence, reflecting long-standing constraints related to land accessibility, regulation, or environmental protection. Rivers display more nuanced, distance-dependent effects, balancing hydrological risks at short distances with amenity or accessibility benefits at intermediate ranges. These findings remain stable across modeling strategies, which reinforces their robustness and interpretability within the covariate set considered here.

The SDDR framework further enhances this analysis by integrating learned spatial representations from local image contexts while preserving an explicit decomposition of covariate-specific partial effects. This hybrid structure enables both improved predictive performance and meaningful interpretation of spatial effects. Incorporating temporal image augmentation by including imagery from preceding time points proved particularly beneficial for small spatial windows, which suggests that local historical context captures relevant short-term dependencies that static models overlook.

The covariate set used throughout this study remains deliberately narrow, and enriching it with transport accessibility, zoning status, or socio-economic indicators is a natural next step. Such an extension would allow the structured component of the SDDR framework to be interpreted in a broader substantive context, while leaving the methodological comparison developed here unchanged.

Despite these advances, the temporal dimension in this study remains limited to a small number of discrete observation dates. A natural and promising extension of this work would be to investigate urbanization as a fully continuous spatio-temporal process by leveraging longer historical map series, including earlier decades where available. Integrating a denser temporal sampling would enable the development of explicit time-series or dynamic SDDR formulations, allowing temporal smooths, lagged effects, or recurrent structures to be incorporated directly into the modeling framework. Such extensions would open new perspectives for understanding long-term urban growth trajectories and for improving the predictive capacity of hybrid spatial models in land-use change analysis.

% =============================================================================
\section*{Acknowledgements}

The research in this project is conducted in the context of the BKG-Gauss Project of the Federal Agency of Cartography and Geodesy (BKG), Frankfurt, Germany. The authors thank the regional geodata authorities of Lower Saxony and Bavaria for access to the land cover data used in this study.

\section*{Disclosure statement}

The authors report there are no competing interests to declare.

\section*{Declaration of generative AI use}

An AI-based language model, ChatGPT (OpenAI), was used to assist with language editing and code refinement during manuscript preparation. No generative AI tool was used to generate data, to perform the statistical analyses, or to produce the scientific content of the article. The authors take full responsibility for the content of the article.

\section*{Funding}

This work was supported by the Federal Agency of Cartography and Geodesy (BKG), Frankfurt, Germany, under Grant No. B12.31-0160121/VV:2.

\section*{Data availability statement}

The geospatial datasets used in this study were provided by the regional geodata authorities of Lower Saxony and Bavaria, namely the Landesamt f\"ur Geoinformation und Landesvermessung Niedersachsen (LGLN) and the Landesamt f\"ur Digitalisierung, Breitband und Vermessung Bayern (LDBV). Since January 1, 2023, several geodata products, including ATKIS\textsuperscript{\textregistered} Basis-DLM and DTK25, have been made available as open data. The datasets used in this study are accessible through the Bavarian open data portal at \url{https://geodaten.bayern.de/opengeodata/OpenDataDetail.html?pn=atkis_basis_dlm}. These data are distributed under the Creative Commons Attribution 4.0 International License (CC BY 4.0), which permits both private and commercial use provided that appropriate attribution is given, a link to the license is included, and any modifications to the data are indicated. The required attribution is Bayerische Vermessungsverwaltung, \url{www.geodaten.bayern.de} (data modified), License CC BY 4.0. The same attribution applies to the ATKIS land use data.

% =============================================================================
\bibliographystyle{tfcad}
\bibliography{references}

\end{document}